\documentclass[reprint, superscriptaddress, nofootinbib]{revtex4-1}

\usepackage{amsmath,amsfonts,amssymb}
\usepackage{appendix}
\usepackage{color}
\usepackage{datetime}
\usepackage{graphicx}
\usepackage{dsfont}
\usepackage{array}
\usepackage[citecolor=blue]{hyperref}
\usepackage{braket}

\newcommand{\mL}{\mathcal{L}}
\newcommand{\mO}{\mathcal{O}}

\newcommand{\pd}{\partial}
\newcommand{\nn}{\nonumber}

\begin{document}

\title{A note on the large charge expansion in 4d CFT}
\author{Gabriel Cuomo}
\email{gabriel.cuomo@epfl.ch}
\affiliation{Theoretical Particle Physics Laboratory (LPTP), Institute of Physics, EPFL, Lausanne, Switzerland}
\date{\today}

\begin{abstract}
In this letter, we discuss certain universal predictions of the large charge expansion in conformal field theories with $U(1)$ symmetry, mainly focusing on four-dimensional theories. We show that, while in three dimensions quantum fluctuations are responsible for the existence of a theory-independent $Q^0$ term in the scaling dimension $\Delta_Q$ of the lightest operator with fixed charge $Q\gg 1$, in four dimensions the same mechanism provides a universal $Q^0\log Q$ correction to $\Delta_Q$. 
Previous works discussing four-dimensional theories failed in identifying this term.
We also compute the first subleading correction to the OPE coefficient corresponding to the insertion of an arbitrary primary operator with small charge $q\ll Q$ in between the minimal energy states with charge $Q$ and $Q+q$, both in three and four dimensions. This contribution does not depend on the operator insertion and, similarly to the quantum effects in $\Delta_Q$, in four dimensions it scales logarithmically with $Q$.
\end{abstract}
\maketitle

\section{Introduction}

In \cite{Hellerman:2015nra} it was argued that, in conformal field theory (CFT), operators with large internal charge can generally  be  associated,  via  the  state-operator  correspondence,  to a superfluid  phase  of  the  theory  on the  cylinder. As a consequence, the universal effective field theory (EFT) description of the hydrodynamic Goldstone modes of the superfluid \cite{Son:2002zn} allows to study systematically correlation functions of these operators \cite{Monin:2016jmo} (see also \cite{Bern1}). The derivative expansion of the EFT coincides with an expansion in inverse powers of the charge. This construction is reviewed in sec. \ref{SecEFT} of this letter for $U(1)$-invariant CFTs.

In the calculation of a given observable within EFT, one may distinguish between classical and quantum contributions. In particular, while the classical contributions depend on a new set of UV-dependent Wilson coefficients at each order in the derivative expansion, the quantum corrections are usually related to the Wilson parameters determining the lower orders; therefore, they are calculable and, in some sense, \emph{universal}. For instance, in the chiral Lagrangian (in the limit of vanishing quark masses) the structure of the $\sim E^2/\Lambda^2$ contribution to pion scattering at center of mass energy $E\ll \Lambda$, $\Lambda$ being the chiral symmetry breaking scale, is fixed by the EFT up to a single Wilson coefficient. The same leading order action determines the non-analytic piece of the one-loop contribution; this is proportional to the logarithm of the cutoff and scales as $(E^2/\Lambda^2)^2\log \Lambda^2/E^2$ \cite{KaplanEFT}. Instead, the finite $(E^2/\Lambda^2)^2$ contributions from the one-loop are renormalized by the higher derivative operators in the chiral Lagrangian; therefore, the analytic $(E^2/\Lambda^2)^2$ term is purely a classical contribution and the EFT predicts its structure only up to a set of new Wilson parameters.

The interplay between classical contributions and the subleading universal quantum corrections obviously plays an important role also in the determination of the CFT data of the theory within the large charge superfluid EFT. The most appreciated example concerns the prediction for the scaling dimension $\Delta_Q$ of the operator with lowest dimension at fixed charge $Q$ in a $U(1)$-invariant three-dimensional CFT. The result is \cite{MoninZeta,Monin:2016jmo}:
\begin{equation}\label{eqIn1}
\Delta_Q\vert_{d=3}=\alpha_1 Q^{\frac{3}{2}}+\alpha_2 Q^{\frac12}-0.0937255+\ldots\,.
\end{equation}
To the first subleading order, the result depends on the value of two Wilson coefficients $\alpha_1$ and $\alpha_2$. The $Q^0$ term corresponds instead to the Casimir energy of the Goldstone mode and it is therefore a one-loop quantum correction. As shown in \cite{Hellerman:2015nra}, its value is fully calculable since it cannot be renormalized by any local counterterm. Remarkably, not only this term does not depend on any new coefficients, but it takes the same universal value for all CFTs whose large charge sector corresponds to a superfluid phase.

The structure of the quantum corrections to $\Delta_Q$ in four-dimensional $U(1)$-invariant CFTs is qualitatively different. Indeed, in that case the scaling dimension $\Delta_Q$ is proportional to $Q^{\frac43}$, with subleading corrections arising from higher derivative terms suppressed by powers of $Q^{\frac23}$. The one-loop Casimir energy scales as $Q^0$ and may therefore be renormalized by the operators associated with the Wilson coefficients contributing to the third order action. Because of this, the authors of \cite{BernSannino1,BernSannino2,Gaume:2020bmp} concluded that the prediction of the large charge EFT in four dimensions concerns only the structure of $\Delta_Q$, but, differently from $d=3$  \eqref{eqIn1}, no fully universal contribution arises in this case; in other words, each order in the $1/Q^{\frac23}$ expansion is proportional to a new independent Wilson coefficient. The main purpose of this letter is to disprove such claim. Indeed, in sec. \ref{SecDeltaQ} we show that the large charge EFT predicts the existence of a calculable and theory-independent $Q^0\log Q$ term in $\Delta_Q$. More precisely, the final result reads:
\begin{equation}\label{eqIn2}
\Delta_Q\vert_{d=4}=\alpha_1Q^{\frac{4}{3}} +\alpha_2 Q^{\frac{2}{3}}
-\frac{1}{48\sqrt{3}}\log Q+
\alpha_3+\ldots\,,
\end{equation}
where the $\alpha_i$'s are UV-dependent Wilson coefficients. The existence of a universal contribution proportional to the logarithm of the charge is fully analogous to the $(E^2/\Lambda^2)^2\log \Lambda^2/E^2$ corrections to pion scattering, and more in general to the standard logarithms of the cutoff scale arising at one-loop in perturbative field theories. As in those cases, the coefficient of the $Q^0$ contribution that is not logarithmically enhanced by the cutoff depends on the precise value of the counterterm renormalizing the quantum loop and it is thus not universal.

As already noticed in \cite{CuomoEpsilon2}, the second subleading quantum correction to $\Delta_Q$, scaling as $Q^{-1}$ ind $d=3$ and as $Q^{-\frac{2}{3}}\log Q$ in $d=4$, is also of some interest, since its value is correlated with the subleading corrections to the  dispersion relation of the  Goldstone. In sec. \ref{SecDeltaQ} we also provide explicit expressions for this correction in both $d=3$ and $d=4$. We also discuss the generalization of eqs. \eqref{eqIn1} and \eqref{eqIn2} to $d=5$ and $d=6$.

Universality is manifest also in other CFT data predicted by the large charge expansion. As a further illustration, in sec. \ref{SecOPE} we consider the correlator of an arbitrary light operator with small $U(1)$ charge $q$ in between the lowest dimensional operators with charge $Q$ and $-(Q+q)$. In \cite{Monin:2016jmo} it was shown that, matching the light operator in terms of the Goldstone field, the EFT predicts that, to leading order, the OPE coefficient scales as $Q^{\frac{\delta}{d-1}}$ times a Wilson parameter that depends on the operator under consideration. Here we show that the first subleading correction to this result is instead independent of the specific operator and, in $d=3$, it is proportional to $q^2Q^{\frac{\delta-1}{2}}$, while in $d=4$ it scales as $q^2Q^{\frac{\delta-2}{3}}\log Q $.

\section{The conformal superfluid EFT}\label{SecEFT}

Consider a $d$ dimensional CFT with $U(1)$ internal symmetry and let us call $\ket{Q}$ the minimal energy state at fixed value of the internal charge for the theory quantized on $\mathbb{R}\times S^{d-1}$ with sphere radius $R$. By the state-operator correspondence $\ket{Q}$ has the same quantum numbers of the operator $\mO_Q(x)$ with lowest dimension at fixed charge $Q$, the energy $E_Q$ being related to the scaling dimension $\Delta_Q$ of the operator as $E_Q=\Delta_Q/R$. Further consider the Euclidean matrix element for an arbitrary number of operator insertions with small quantum numbers in between the ground state $\ket{Q}$:
\begin{equation}\label{eq_ch21_PI_cyl}
\braket{Q,\tau_{out}|\mO_m(\tau_m,\hat{n}_m)\ldots\mO_1(\tau_1,\hat{n}_1)|Q,\tau_{in}}\,,
\end{equation}
where $\tau=R\,\log\left(|x|/R\right)$ denotes Euclidean time on the cylinder, $\hat{n}^\mu=x^\mu/|x|$ specify the coordinates on the sphere and the states are defined in Schr\"odinger picture,
\begin{align}
&\ket{Q,\tau_{in}}\equiv e^{H\tau_{in}}\ket{Q}=e^{E_{Q}\tau_{in}}\ket{Q}\,,\\
&\bra{Q,\tau_{out}}\equiv\bra{Q}e^{-H\tau_{out}}
=\bra{Q}e^{-E_{Q}\tau_{out}}\,.
\end{align}

The basic observation of \cite{Hellerman:2015nra,Monin:2016jmo} can be phrased as follows. In the limit $Q\gg 1$ we expect the path-integral describing the matrix element \eqref{eq_ch21_PI_cyl} to be dominated by  semiclassical saddle-point trajectories  characterized  by  a  specific pattern  of  symmetry  breaking. The most natural situation corresponds to the case in which the leading trajectory is characterized by a superfluid pattern \cite{Monin:2016jmo}, which is defined as, in obvious notation,
\begin{equation}\label{eq_ch21_SSB_U(1)}
SO(d+1,1)\times U(1)\longrightarrow SO(d)\times\bar{D}\,,
\end{equation}
where $SO(d+1,1)$ is the $d$-dimensional conformal group, $SO(d)$ the Euclidean rotation group, $\widehat{Q}$ is the $U(1)$ generator and $\bar{D}=D+\mu R\, \widehat{Q}$ is a linear combination of the dilation generator $D$, corresponding to the cylinder Hamiltonian, and the internal generator $\widehat{Q}$. Here $\mu$ defines the chemical potential. Under the assumption that the leading semiclassical trajectory realizes the pattern \eqref{eq_ch21_SSB_U(1)}, the properties of the ground state and its fluctuations are characterized by the corresponding Goldstone excitations. When not implied differently by additional symmetries, such as supersymmetry \cite{LargeQSUSY1}, additional degrees of freedom are expected to be separated by a finite gap from the Goldstones and may be integrated out. We can therefore \emph{effectively} compute the path-integral corresponding to the matrix element \eqref{eq_ch21_PI_cyl} using a low energy action for the Goldstone degrees of freedom.~\footnote{Notice that despite the leading trajectory in the path-integral induces the symmetry breaking pattern in eq. \eqref{eq_ch21_SSB_U(1)}, the state $\ket{Q}$ does not break the $U(1)$ symmetry; see \cite{Monin:2016jmo} for additional comments on this point.}

The pattern \eqref{eq_ch21_SSB_U(1)} may be realized in terms of a single shift-invariant superfluid Goldstone $\chi(x)=-i\mu\tau+\pi(x)$ \cite{Son:2002zn,Monin:2016jmo}, associated to the breaking of the $U(1)$ symmetry, where the chemical potential $\mu$ will be determined eventually by the charge $Q$. The action is easily constructed in an expansion in derivatives over $\mu$ noticing that the following \emph{modified metric} $\hat{g}_{\mu\nu}=g_{\mu\nu}(\pd\chi)^2$, where $(\pd\chi)=(-\pd_\mu\chi \,g^{\mu\nu}\pd_\nu\chi)^{1/2}$, is invariant under Weyl transformations. Denoting $\hat{\nabla}_\mu$ and $\hat{\mathcal{R}}^\rho_{\;\mu\sigma\nu}$ the covariant derivative and the Riemann tensor obtained from $\hat{g}_{\mu\nu}$ and discarding terms which vanish on the equations of motion of the leading order Lagrangian \cite{Weinberg2}, the action reads
\begin{equation}\label{eq_ch21_LagNLO}
S=S^{(1)}+S^{(2)}+S^{(3)}+\ldots\,
\end{equation}
\vspace*{-0.6cm}
\begin{align}
S^{(1)}=&-c_1\int d^dx\sqrt{\hat{g}} =-c_1\int d^dx\sqrt{g} (\pd\chi)^d\,,\\
\nonumber
S^{(2)} =&\int d^dx\sqrt{\hat{g}}\left\{c_2\hat{\mathcal{R}}
-c_3\hat{\mathcal{R}}^{\mu\nu}\pd_\mu\chi\pd_\nu\chi\right\}\\  
\nonumber
=&c_2\int d^dx\sqrt{g} (\pd\chi)^d\left\{ \frac{\mathcal{R}}{(\pd\chi)^2}
+\ldots\right\}\\
\label{eq_ch21_LagNLO2}
-&c_3\int d^dx\sqrt{g} (\pd\chi)^d\left\{\mathcal{R}_{\mu\nu}\frac{\pd^\mu\chi\pd^\nu\chi}{(\pd\chi)^4}
+\ldots
\right\}\,,
\\ \nn
S^{(3)} =&c_4\int d^dx\sqrt{\hat{g}}\hat{\mathcal{R}}^2+\ldots\\
=&c_4\int d^dx\sqrt{g} (\pd\chi)^d\left\{\frac{\mathcal{R}^2}{(\pd\chi)^4}+\ldots\right\}+\ldots\,, \label{eq_ch21_LagNLO3}
\end{align}
where $\mathcal{R}^\rho_{\;\mu\sigma\nu}$ is the Riemann tensor deriving from the cylinder metric $g_{\mu\nu}$ and the dots stand for terms with at least two covariant derivatives acting on $\pd_\mu\chi$; their precise form can be found using the standard formulas relating curvature invariants of two Weyl equivalent metrics reported, e.g., in \cite{Wald}. The $c_i$'s are Wilson coefficients, whose value depends on the microscopic dynamics of the specific underlying CFT. In the simplest scenario, corresponding to an underlying strongly coupled theory, the $c_i$'s are given by inverse powers of $4\pi$'s according to generalized dimensional analysis \cite{Georgi:1992dw}. Weakly coupled theories correspond instead to non-generic sizes for the Wilson coefficients, see e.g. \cite{CuomoEpsilon2} for an explicit example. For future convenience, we also wrote the only term of the third order action which does not vanish on the background $\pd_\mu\chi=-i\mu\delta_\mu^0$ in $d=4$. 

Let us prove that no other terms at order $\mu^{d-4}\vert_{d=4}=\mu^0$ contribute on the background solution in $d=4$. To this aim, notice that $\nabla_\mu\pd_\nu\chi=0$ and that $\mathcal{R}^0_{\;\mu\nu\rho}=\mathcal{R}^\mu_{\;0\nu\rho}=
\mathcal{R}^\mu_{\;\nu0\rho}=\mathcal{R}^\mu_{\;\nu\rho0}=0$, so that any contraction of the Riemann tensor with $\pd_\mu\chi$ vanishes. Therefore, the only two additional invariants that need to be considered in $d=4$ are $\hat{W}_{\mu\nu\rho\sigma}\hat{W}^{\mu\nu\rho\sigma}$ and $\hat{E}$, where $W_{\mu\nu\rho\sigma}$ and $E$ are, respectively, the Weyl tensor and the Gauss-Bonnet term \cite{Wald}. However both of them vanish identically on the background. Indeed the first is Weyl invariant and the metric $\hat{g}_{\mu\nu}$ is conformally equivalent to flat space, while the second one in four dimensions coincides with the Euler density, whose integral is a topological invariant and vanishes on the cylinder. Finally, due to the Weyl anomaly in four dimensions, at the same order the effective action must include also the following Wess-Zumino term \cite{Komargodski:2011vj,Hellerman:2015nra}:
\begin{multline}
S_{WZ}\vert_{d=4}=\int d^4x\sqrt{g}\log(\pd\chi)\left[-aE\right.\\
\left.+cW_{\mu\nu\rho\sigma}W^{\mu\nu\rho\sigma}\right]+\ldots\,,
\end{multline}
where $a$ and $c$ are the trace anomalies and the dots stand again for terms
with at least two covariant derivatives acting on $\pd_\mu\chi$. Also this term vanishes on the superfluid solution by considerations similar to the ones above.



\section{The scaling dimension}\label{SecDeltaQ}

Using the action \eqref{eq_ch21_LagNLO}, in this section we extract the scaling dimension of the lightest operator with fixed charge $Q\gg 1$. We shall present the calculation in an arbitrary number spacetime of spacetime dimensions $d$. This will allow us to immediately identify the main differences between even and odd spacetime dimensions. Furthermore, since the previously derived action is Weyl and $U(1)$-invariant in an arbitrary number of spacetime dimensions, working for arbitrary $d$ provides us with a natural regulator for the quantum corrections to the energy.

Following \cite{Monin:2016jmo}, we can compute $E_Q$ considering the Euclidean evolution amplitude of an arbitrary charge $Q$ state $\ket{Q,X}$:
\begin{equation}
\braket{Q,X|e^{-HT}|Q,X}\propto e^{-E_QT}\,,
\end{equation}
where we used that in the limit $T\rightarrow\infty$ any state with charge $Q$ will project to the ground state. A convenient choice of the state leads to the following path-integral \cite{Monin:2016jmo}
\begin{multline}\label{eq_ch21_PI_U(1)}
\braket{Q|e^{-HT}|Q}\propto
\int \mathcal{D}\chi_i\mathcal{D}\chi_f\psi_Q(\chi_i)\psi_Q^*(\chi_f)\\
\times
\int_{\chi=\chi_i}^{\chi=\chi_f}\mathcal{D}\chi\exp\left[-\int_{-T/2}^{T/2} d\tau\int d^{d-1}x\sqrt{g}\mathcal{L}\right]\,,
\end{multline}
where the wave-functionals $\psi_Q(\chi)$ ensure that the initial and final state have the correct $U(1)$ charge:
\begin{equation}
\psi_Q(\chi)=\exp\left[\frac{i\,Q}{R^{d-1}\Omega_{d-1}}\int d^{d-1}x\sqrt{g}\chi\right].
\end{equation}
Here $\Omega_{d-1}=\frac{2 \pi^{d/2}}{\Gamma(d/2)}$ is the volume of the $d-1$ dimensional sphere. 

For large $Q$, the integral \eqref{eq_ch21_PI_U(1)} can be computed semiclassically around the saddle-point solution
\begin{equation}\label{eq_ch21_sol}
\chi=-i\mu \tau+\pi_0\,.
\end{equation}
Notice that on the solution the term proportional to $c_3$ in the action \eqref{eq_ch21_LagNLO2} vanishes, since $\mathcal{R}_{00}=0$ on $\mathbb{R}\times S^{d-1}$. Here $\pi_0$ is an integration constant and the field is analytically continued away from the real axis. The variation of the field at the boundary fixes the chemical potential $\mu$ in terms of the charge $Q$ via 
\begin{align}\label{eq_ch21_JO1}
i\frac{\pd\mL}{\pd\dot{\chi}}=\frac{Q}{R^{d-1}\Omega_{d-1}}\,.
\end{align}
Solving this equation perturbatively for large $Q$, we find
\begin{gather}\nn
R\mu=\widetilde{Q}^\frac{1}{d-1}\left[1+
\frac{c_2  (d-2)^2 }{c_1d  }
\widetilde{Q}^{-\frac{2}{d-1}}+\mO\left(\widetilde{Q}^{-\frac{4}{d-1}}\right)
\right]\,,\\ \label{eq_ch21_mu}
\widetilde{Q}\equiv\frac{Q}{c_1 d\,\Omega_{d-1}}\,.
\end{gather}
For $Q\gg 1$ we thus have $\mu\propto Q^{\frac{1}{d-1}}$, with subleading corrections suppressed by powers of $Q^{-\frac{2}{d-1}}$.~\footnote{Notice however that for $c_1\gg 1$, as it is expected in weakly coupled theories, the chemical potential may be parametrically smaller than $Q^{\frac{1}{d-1}}$.} Computing the action on this solution, we find the classical contribution to the energy of the state:
\begin{equation}\label{eq_DeltaQ_U(1)_classical}
\Delta_Q\vert_{classical}=\alpha_1 Q^{\frac{d}{d-1}}+\alpha_2 Q^{\frac{d-2}{d-1}}+
\alpha_3 Q^{\frac{d-4}{d-1}}+\ldots
\,,
\end{equation}
where the $\alpha_i$'s are combination of the Wilson coefficients. For instance, the first two read:
\begin{equation}\label{eq_ch21_alpha}
\alpha_1= \frac{ c_1(d-1)\Omega_{d-1} }{\left(c_1 d\, \Omega_{d-1} \right)^{\frac{d}{d-1}}}\,,\quad
\alpha_2=\frac{c_2 (d-1)(d-2) \Omega_{d-1}}{ 
\left(c_1 d \,\Omega _{d-1}\right)^{\frac{d-2}{d-1}}}\,.
\end{equation}
The scaling with $Q$ of the leading term in eq. \eqref{eq_DeltaQ_U(1)_classical} could have been inferred on dimensional grounds \cite{Hellerman:2015nra}. Indeed, for a scale invariant theory in the semiclassical regime the charge density $\rho$ and the energy density $\varepsilon$ are expected to obey a local relation of the form $\varepsilon\propto \rho^{\frac{d}{d-1}}$. Subleading terms are suppressed by the ratio of the cutoff and the compactification scale $(R^{-1}/\mu)^2\sim Q^{\,-\frac{2}{d-1}}$; this structure follows from the fact that the EFT action depends analytically on the curvature invariants. 

We now want to consider quantum corrections to eq. \eqref{eq_DeltaQ_U(1)_classical}. To this aim, we define $\chi(x)=-i\mu \tau+\pi(x)$ and we expand the low energy action to quadratic order in the fluctuations:
\begin{multline}\label{eq_ch21_deltaL}
S\simeq \frac{ d(d-1)}{2}c_1\mu^{d-2}\\
\times\int d^dx\sqrt{g}\left[\dot{\pi}^2+\frac{1}{d-1}(\pd_i\pi)^2+\mO\left(\nabla^4/\mu^2\right)\right]\,.
\end{multline}
This action describes a phonon mode with speed of sound $c_s^2=\frac{1}{d-1}$, as it is mandated by tracelesness of the energy momentum tensor. More precisely, upon including the subleading correction to the quadratic phonon action, we find that the dispersion relation reads
\begin{gather}\nn
\omega_{\ell}=\frac{1}{\sqrt{d-1}} J_{\ell}
+\frac{\gamma}{Q^{\frac{2}{d-1}}}\left(\frac{J^3_{\ell}R^2}{d-1}-J_{\ell}\right)
+\mO\left(\frac{J^5_\ell R^4}{Q^{\frac{4}{d-1}}}\right)\,,
\\
\label{eq_spectrum_phonons}
\gamma=\frac{\left[c_2 (d-2)+c_3\right](d-2)}{c_1^{\frac{d-3}{d-1}} d\sqrt{d-1} \left( d\,\Omega_{d-1}\right)^{\frac{-2}{d-1}}}\,,
\end{gather}
where $J_{\ell}^2=\ell(\ell+d-2)/R^2$ is the $\ell$th eigenvalue of the Laplacian on the sphere. The Fock space of these modes, except  for the zero mode which relates different charge sectors, describes operators with the same $U(1)$ charge $Q$ but with higher scaling dimension: $\Delta=\Delta_Q+\sum n_{\ell}R\omega_{\ell}$. In particular, the descendants correspond to states involving a number $q>0$ of spin  one quanta, each increasing  the energy by $\omega_1 = 1/R$.

The one-loop contribution to the energy is given by the fluctuation determinant arising from the Gaussian integration of eq. \eqref{eq_ch21_deltaL}:
\begin{align}\nn
\frac{T}{R}\delta\Delta_Q^{(1)} &=\frac{1}{2}\log\det\left[-\pd_\tau^2-\frac{1}{d-1}\nabla^2+\mO\left(\nabla^4/\mu^2\right)\right]
\end{align}
where $\nabla^2=|g^{ij}|\nabla_i\nabla_j$ is the Laplacian on the $d-1$ dimensional sphere. Proceeding as in \cite{CuomoEpsilon1}, one can show that the value of $\delta\Delta_Q^{(1)}$ in dimensional regularization coincides with a sum of zero point phonon energies:
\begin{equation}\label{eq_ch21_DeltaQ_1loop}
\delta\Delta_Q^{(1)}=\frac{1}{2}\sum_{\ell}n_{\ell,d}R\omega_{\ell}=
\beta_0+\beta_1Q^{-\frac{2}{d-1}}+\ldots\,.
\end{equation}
Here $n_{\ell,d}=\frac{(2\ell+d-2)\Gamma(\ell+d-2)}{\Gamma(\ell+1)\Gamma(d-1)}$ is the multiplicity of the Laplacian eigenvalue $J_{\ell}$ on a $d-1$ dimensional sphere. We formally wrote the result in a large $Q$ expansion in terms of dimensionless coefficients $\beta_i$'s, whose specific value depends on the number of dimensions $d$. Notice that $\beta_0$ cannot depend on the $c_i$'s, because the sound-speed in eq. \eqref{eq_ch21_deltaL} is fixed by conformal invariance at leading order in $Q$. Summing the quantum corrections to the classical result \eqref{eq_DeltaQ_U(1)_classical} we find
\begin{align}\nn
\Delta_Q=Q^{\frac{d}{d-1}}&\left[\alpha_1 +\alpha_2 Q^{\,-\frac{2}{d-1}}+
\alpha_3 Q^{\,-\frac{4}{d-1}}+\ldots\right]\\
+Q^0&\left[\beta_0+\beta_1 Q^{\,-\frac{2}{d-1}}+\ldots\right]+
\ldots\,. \label{eq_DeltaQ_U(1)}
\end{align}
We neglected the two-loop correction to the energy, which scales as $\sim Q^{\,-\frac{d}{d-1}}$ (up to logarithms of $Q$).

Eq. \eqref{eq_DeltaQ_U(1)} immediately shows the main difference between even and odd $d$. Indeed, the contribution from the classical solution, associated to the coefficients $\alpha_i$, does not contain any term scaling as $Q^0$ for non-even $d$. This implies that, in odd spacetime dimensions, the one-loop correction \eqref{eq_ch21_DeltaQ_1loop} cannot be renormalized by any local counterterm and it is hence finite and calculable. In particular, since $\beta_0$ is independent of the Wilson coefficients, the $Q^0$ contribution takes the same \emph{universal} value for all three-dimensional $U(1)$-invariant CFTs whose large charge sector is described by a superfluid phase. 
The explicit result for the quantum corrections in $d=3$ can be found proceeding as in the appendix of \cite{CuomoEpsilon1} and reads~\footnote{The value of $\beta_1$ here corrects eqs. (36) and (38) of \cite{CuomoEpsilon2}, where the piece proportional to $c_3\vert_{here}=\lambda\alpha_2\vert_{there}$ was incorrect; this mistake however did not affect the main results of that work, where the authors considered a specific UV complete theory such that $c_3\vert_{here}=\lambda\alpha_2\vert_{there}=0$ in the EFT to tree-level accuracy.}
\begin{align}\nn
\beta_0\vert_{d=3}&=-0.0937255\,,\\ \label{eq_quantum_Q0}
\beta_1\vert_{d=3}&=(c_2+c_3)\times 1.21666\,.
\end{align}
This result for $\beta_0$ is in agreement with the value originally derived in \cite{MoninZeta} within zeta-function regularization.
Summing everything, the final result in $d=3$ reads:
\begin{gather}\nn
\Delta_Q\vert_{d=3}=\alpha_1 Q^{\frac32}+\alpha_2Q^{\frac12}-0.0937255+\alpha_3 Q^{-\frac12}
\\+(c_2+c_3)\times 1.21666\times Q^{-1}+\mO\left(Q^{-\frac32}\right)\,.
\label{DeltaQ3}
\end{gather} 
Notice that the combination $c_2+c_3$ controlling the $Q^{-1}$ contribution can be extracted from the first subleading correction to the dispersion relation of the Goldstone mode \eqref{eq_spectrum_phonons}. A non-trivial check of both this relation and the value of the $Q^0$ term in a specific three-dimensional weakly coupled model was provided in \cite{CuomoEpsilon2}. The value of the $Q^0$ term was previously verified in \cite{Anton} at large $N$ for monopole operators and it is in agreement with the result of Monte-Carlo simulations \cite{Banerjee:2017fcx}.

Conversely, in $d=4$ the $\beta_0$ term in eq. \eqref{eq_DeltaQ_U(1)} can be renormalized by the classical contribution proportional to $\alpha_3$, and it is hence expected to be divergent. Similarly for $\beta_1$ that may be renormalized by $\alpha_4$. Indeed, 
we find that their expressions within dimensional regularization contain a pole for $d\rightarrow 4$: 
\begin{align}\nn
\beta_0\vert_{d\rightarrow 4}&=\frac{1}{16\sqrt{3}(d-4)}+\text{finite}\,,
\\
\beta_1\vert_{d\rightarrow 4}&=-
\frac{7 \pi ^{\frac43} (2 c_2+c_3) }{48\sqrt{3}\, c_1^{1/3}(d-4) }
+\text{finite}\,.
\end{align}
The finite parts can always be re-absorbed in the definition of the Wilson coefficients of the operators contributing to the third and higher orders in the action and are hence irrelevant for our purposes.

As typical in quantum field theory, the divergent part of a quantum loop is related to a calculable logarithm of the UV scale. To see this mechanism at work here, notice that the UV divergence associated to $\beta_0$ can be reabsorbed in the definition of the bare coefficient $c_4$ upon writing it as
\begin{equation}
c_4=-\frac{1}{36\Omega_3}\times\frac{1}{16\sqrt{3}(d-4)}+c_4^{ren.}\,,
\end{equation}
where $c_4^{ren.}$ is finite and we used $\mathcal{R}^2=36/R^4$ in $d=4$ to obtain the prefactor. Adding the contribution from eq. \eqref{eq_ch21_LagNLO3} to $\beta_0$ and expanding the result for $d\rightarrow 4$ we find:
\begin{align}\nn
\beta_0+\frac{R}{T}S^{(3)}&= \lim_{d\rightarrow 4}\left[\frac{1- (R\mu)^{d-4}}{16\sqrt{3}(d-4)}\right]+\text{finite}\times Q^0\\ \label{LogQ0}
&=-\frac{1}{16\sqrt{3}}\log R\mu+\text{finite}\times Q^0\,,
\end{align}
which indeed contains a logarithm of $\mu\sim Q^{1/3}$. The divergent contribution from $\beta_1$ may be similarly renormalized by higher order terms, schematically of the form $\sqrt{\hat{g}}\hat{\mathcal{R}}^3\sim(\pd\chi)^{d-6}\mathcal{R}^6$, their precise expression being irrelevant. Proceeding as before, we find a calculable contribution of the form $(R\mu)^{-\frac23}\log R\mu$. Eventually, using eq. \eqref{eq_ch21_mu} to relate $\mu$ and $Q$, the final result reads:
\begin{equation}
\begin{gathered}
\left.\Delta_Q\right\vert_{d=4}=\alpha_1Q^{\frac{4}{3}} +\alpha_2 Q^{\frac{2}{3}}
-\frac{1}{48\sqrt{3}}\log Q+
\alpha_3 \\
+
\frac{7 \pi ^{\frac43} (2 c_2+c_3) }{144 \sqrt{3}\, c_1^{1/3} }
Q^{\,-\frac{2}{3}}\log Q 
+\alpha_4 Q^{\,-\frac{2}{3}}+
\mO\left(Q^{\,-\frac{4}{3}}\right)\,.
\label{eq_DeltaQ_U(1)4}
\end{gathered}
\end{equation}

Eq. \eqref{eq_DeltaQ_U(1)4} is the main result of this letter. It shows that the large charge expansion in $d=4$ predicts the existence of a universal and calculable $Q^0\log Q$ contribution to the energy. Some previous works studying four-dimensional models \cite{BernSannino1,BernSannino2,Gaume:2020bmp} have failed in identifying the existence of such term, incorrectly claiming instead that $\Delta_Q$ does not contain any theory-independent contribution in $d=4$. 
Perhaps, this mistake was induced by the use of an arguably less transparent regularization scheme for the quantum corrections, namely zeta-function regularization, which breaks scale invariance at intermediate steps.
In our approach, the result \eqref{eq_DeltaQ_U(1)4} was instead obtained in a straightforward manner within dimensional regularization, which preserves both the conformal and the internal symmetry at every step of the calculation. 
Notice also that, similarly to the $Q^{-1}$ contribution in eq. \eqref{DeltaQ3}, the coefficient of the  $Q^{\,-\frac{2}{3}}\log Q$ term is related to the subleading correction to the dispersion relation of the phonon \eqref{eq_spectrum_phonons}. 

The result \eqref{eq_DeltaQ_U(1)4} applies to any $4d$ CFT with $U(1)$ symmetry whose large charge sector is a superfluid, including superconformal theories in which supersymmetry is fully broken at large charge \cite{Hellerman:2015nra}. In the future, it should be possible to verify explicitly the value of the universal logarithmic contributions in perturbative models, along the lines of \cite{CuomoEpsilon2,Anton}. 

Finally, we mention that our results admit obvious extensions to higher spacetime dimensions. For completeness, we provide the expression of $\Delta_Q$ both in $d=5$ and $d=6$ to order $Q^0$:
\begin{multline}
\Delta_Q\vert_{d=5}=\alpha_1 Q^{5/4}+\alpha_2 Q^{3/4}+\alpha_3 Q^{1/4}\\
-0.1079+\mO\left(Q^{-1/4}\right) \,,
\end{multline}
\begin{multline}
\Delta_Q\vert_{d=6}=\alpha_1 Q^{6/5}+\alpha_2 Q^{4/5}+\alpha_3 Q^{2/5}\\
-\frac{1}{60 \sqrt{5}}\log Q
+\alpha_4 Q^{0}+\mO\left(Q^{-2/5}\right)\,.
\end{multline}

The superfluid description at large charge might apply in some two dimensional models as well. That case however is special, since the two-dimensional cylinder is a flat manifold and the EFT coincides with the theory of a free compact boson to leading order. In particular, the $Q^0$ Casimir energy of the Goldstone boson is not renormalized by any classical term differently from higher even dimensions. In the future, it would be interesting to identify explicitly interacting $2d$ CFTs whose large charge sector is described by the superfluid EFT.
%

\section{OPE coefficients}\label{SecOPE}

Suppose that in the CFT under consideration there exists a scalar primary operator with scaling dimension $\delta\ll \Delta_Q^{1/2}$ \footnote{This ensures that the terms arising from the expansion of the term $(\pd\chi)^{\delta}$ of eq. \eqref{eq_match_scalar} into canonically normalized field fluctuations are small.} and small charge $q$, in a sense that will be quantified later. We can reconstruct this operator in the large charge EFT by matching its quantum numbers in terms of the Goldstone field as \cite{Monin:2016jmo}
\begin{equation} \label{eq_match_scalar}
\mO^{(\delta)}_q =C_{\delta,q}^{(1)}(\pd\chi)^{\delta}e^{iq\chi}-C_{\delta,q}^{(2)}(\pd\chi)^{\delta-2}\left[\mathcal{R}+\ldots\right]e^{iq\chi}+\ldots
\,.
\end{equation}
As for the $c_i$'s in the effective action \eqref{eq_ch21_LagNLO}, $C_{\delta,q}^{(1)}$ and $C_{\delta,q}^{(2)}$ are $Q$-independent Wilson coefficients whose value is not predicted by the EFT.

We can use the expression \eqref{eq_match_scalar} for the operator to compute  the matrix element $\lambda_{q}^{(\delta)}\equiv\braket{Q+q|\mO^{(\delta)}_q|Q}$ in $d=3$ and $d=4$. To this aim, we insert the expression \eqref{eq_match_scalar} in the path-integral \eqref{eq_ch21_PI_cyl}
\begin{multline}\label{eq_ch21_3pts2}
\braket{Q+q,\tau_{out}|\mO^{(\delta)}_q(x_c)|Q,\tau_{in}}\\
=\int\mathcal{D}\chi
\left[C_{\delta,q}^{(1)}\left(\pd\chi\right)^\delta+\ldots\right]
e^{-S_{mod}}
\,,
\end{multline}
where we included the $e^{iq\chi}$ contribution from the operator insertions and the wave-functions in the definition of the following modified action
\begin{equation}\label{eq_ch21_Smod3}
\begin{split}
S_{mod}[\chi]&=S[\chi]+i\frac{Q+q}{\Omega_{d-1}}\int d\Omega_{d-1}\chi_f\\
&-iq\chi(\tau_c,\hat{n}_c)-i\frac{Q}{\Omega_{d-1}}\int d\Omega_{d-1}\chi_i\,.
\end{split}
\end{equation}
To compute the path-integral in a saddle-point approximation, we look for a solution of the equations of motion of the action $S_{mod}$ to leading order in derivatives:
\begin{equation}\label{eq_ch21_3ptEq}
\nabla_\mu j^\mu(x)=q\delta^{(d)}(x-x_c)\,,
\end{equation}
where $\delta^{(d)}(x-x_c)=\delta(\tau-\tau_c)\delta^{d-1}(\hat{n}-\hat{n}_c)/\sqrt{g}$ and $j^\mu=i\pd\mathcal{L}/\pd(\pd_\mu\chi)$ is the Noether current. In the limit $T\rightarrow\infty$ the boundary conditions read
\begin{align}\nn
j^\mu(x)\xrightarrow{\tau\rightarrow-\infty}\delta^\mu_0\frac{Q}{R^{d-1}\Omega_{d-1}}\,,\\
 j^\mu(x)\xrightarrow{\tau\rightarrow+\infty}\delta^\mu_0\frac{Q+q}{R^{d-1}\Omega_{d-1}}\,.
\end{align}
Physically, we can think of equation \eqref{eq_ch21_3ptEq} as a non-linear version of the electrostatic Gauss-law (where the current is not an exact form), the scalar operator acting as a point-like source with charge $q$, slightly deforming the path-integral. For sufficiently small $q$, we may solve this equation expanding the field around the superfluid solution. To leading order the solution coincides with \eqref{eq_ch21_sol} and we find that the path-integral evaluates to
\begin{multline}
C^{(1)}_{\delta,q}\mu^{\delta}
e^{-\Delta_Q(\tau_{out}-\tau_{in})-q\mu (\tau_{out}-\tau_{c}) }\\ 
\label{eq_ch21_3ptLO}
\approx C^{(1)}_{\delta,q}\mu^{\delta} e^{-\Delta_{Q+q}(\tau_{out}-\tau)-\Delta_Q (\tau-\tau_{i})} \,,
\end{multline}
where in the second line we used
\begin{equation}\label{eq_mu_pd}
qR\mu=q\frac{\pd \Delta_Q}{\pd Q}\approx\Delta_{Q+q}-\Delta_Q\,.
\end{equation}
Using \eqref{eq_ch21_mu} we find that the EFT structure predicts the following scaling law for the OPE coefficient:
\begin{equation}\label{eq_ch21_3pt0}
\lambda_{q}^{(\delta)}\propto  Q^{\frac{\delta}{d-1}}\,.
\end{equation}
This result was originally presented in \cite{Monin:2016jmo}.

We now proceed to extend this analysis to the next order by expanding the solution in fluctuations $\pi(x)=\chi(x)+i\mu\tau-\pi_0$. The linearized problem reads:
\begin{equation}\label{eqAppCFT3pteqLin}
i c_1d(d-1)\mu^{d-2}\left(\pd_{\tau}^2+\frac{1}{d-1}\nabla^2\right)\pi(x)=
q\frac{\delta^{(d)}(x-x_c)}{\sqrt{g}}\,.
\end{equation}
This equation can be straightforwardly solved expanding the field into Gegenbauer polynomials:
\begin{align}\nn
\pi(x) &=i\frac{q/(R^{d-1}\Omega_{d-1})}{c_1d(d-1)\mu^{d-2}}\left[
\vphantom{\sum_{\ell=1}^{\infty}\frac{2\ell+d-2}{d-2}
\frac{e^{-\omega_{\ell}|\tau-\tau_c|}}{2\omega_{\ell}}
C_{\ell}^{\left(\frac{d}{2}-1\right)}\left(\hat{n}\cdot\hat{n}_c\right)}
-(\tau-\tau_c)\theta(\tau-\tau_c)\right.\\ 
&\left.+
\sum_{\ell=1}^{\infty}\frac{2\ell+d-2}{d-2}
\frac{e^{-\omega_{\ell}|\tau-\tau_c|}}{2\omega_{\ell}}
C_{\ell}^{\left(\frac{d}{2}-1\right)}\left(\hat{n}\cdot\hat{n}_c\right)\right]\,. \label{eqAppCFT3ptSol}
\end{align}
Using $\mu\sim Q^{\frac{1}{d-1}}$ this implies that the field scales as $\pi(x)\sim q/Q^{\frac{d-2}{d-1}}$. Corrections arising from nonlinear terms in the expansion are suppressed by a relative power of $q/Q$ with respect to the leading expression \eqref{eqAppCFT3ptSol}. 
Plugging this solution in the action \eqref{eq_ch21_Smod3} and extracting the coordinate dependence as before, we express the OPE coefficient as
\begin{multline}\label{eq_OPE_threept_U(1)}
\lambda_{q}^{(\delta)}=C^{(1)}_{(\delta,q)}(R\mu)^{\delta}\left[1+i\frac{q}{2}
\pi(x_c)+\ldots\right]\\+C^{(2)}_{(\delta,q)}(R\mu)^{\delta-2}(d-1)(d-2)+\ldots
\end{multline}
From eq. \eqref{eq_OPE_threept_U(1)} we infer that the modification of the superfluid profile induces a relative correction to the OPE coefficient proportional to $q\pi(x_c)\sim q^2/Q^{\frac{d-2}{d-1}}$. In order for this term to be subleading we assume $q^2\ll Q^{\frac{d-2}{d-1}}$.  Similarly to the discussion below eq. \eqref{eq_DeltaQ_U(1)}, this scale coincides with some integer power of the one controlling the derivative expansion, given by $Q^{-\frac{2}{d-1}}$, only in even dimensions. As before, this implies a different structure in the predictions in $d=3$ and $d=4$. Notice indeed that the value of $\pi(x_c)$ is finite only for sufficiently negative $d$; thus, in general, its value is obtained by analytic continuation in $d$ and, similarly to the Casimir energy \eqref{eq_ch21_DeltaQ_1loop} in $d=4$, may contain poles for integer values of $d$. We shall discuss the case of a three-dimensional and that of a four-dimensional theory separately in the following.  

Consider first $d=3$. In this case, the first subleading correction to eq. \eqref{eq_ch21_3pt0} cannot be renormalized by the operator-dependent coefficients of the matching \eqref{eq_match_scalar}. Correspondingly, evaluating the value of $\pi(x_c)$ within dimensional regularization we find a finite result. Absorbing all the anyway unknown constants in a new Wilson parameter $\eta^{(1)}_{(\delta,q)}$, the OPE coefficient reads
\begin{multline}\label{eqOPE3d}
\lambda_{q}^{(\delta)}\big\vert_{d=3}= Q^{\delta/2}\left[\eta^{(1)}_{(\delta,q)}\left(1+
0.05051\times\frac{q^2}{\sqrt{c_1Q}}\right)\right.
\\
\left.
+\mO\left(Q^{-1}\right)\vphantom{\left(1+
0.05051\times\frac{q^2}{\sqrt{c_1Q}}\right)}\right]\,.
\end{multline}
The second term in round brackets provides the first correction to the leading order result.~\footnote{This term was neglected in the bootstrap analysis of \cite{Jafferis:2017zna}, but it can be easily checked to be consistent with the crossing conditions discussed there.} Its coefficient is entirely fixed in terms the same parameter $c_1$ controlling the scaling dimension $\Delta_Q$ at leading order (see eqs. \eqref{eq_DeltaQ_U(1)_classical} and \eqref{eq_ch21_alpha}) and it is hence independent of the specific operator under consideration. 

This situation is to be contrasted with $d=4$, in which case the corrections arising from the modification of the profile \eqref{eq_ch21_sol} are renormalized from the first subleading term in the operator matching \eqref{eq_match_scalar}, proportional to $C^{(2)}_{(\delta,q)}$. Proceeding as we did above eq. \eqref{eq_DeltaQ_U(1)4}, we find that $\pi(x_c)$ has a pole for $d\rightarrow 4$. This implies that there exists a calculable logarithmic correction which is independent of $C^{(2)}_{(\delta,q)}$. The final result reads
\begin{multline}\label{eqOPE4d}
\lambda_{q}^{(\delta)}\big\vert_{d=4}= Q^{\delta/3}\left[\eta^{(1)}_{(\delta,q)}\left(1-\frac{q^2 Q^{-2/3}\log Q}{48 \sqrt{3} \pi ^{2/3} c_1^{1/3}}
\right)\right.\\
\left.+
\eta^{(2)}_{(\delta,q)} Q^{-2/3}+
\mO\left( Q^{-4/3}\right)\vphantom{\frac{q^2 Q^{-2/3}}{24 \sqrt{3} \pi ^{2/3} c_1^{1/3}}}\right]\,,
\end{multline}
where the $\eta^{(i)}_{(\delta,q)}$'s are independent Wilson coefficients . The coefficient of the $Q^{-2/3}\log Q$ term in round brackets does not depend on the specific operator under consideration. 

Finally, we remark that the results \eqref{eqOPE3d} and \eqref{eqOPE4d} hold also for primary operators in spin $\ell$ traceless-symmetric representations. To see this, we notice that a spin $\ell$ primary, with scaling dimension $\delta$ and charge $q$, can be matched in the low energy EFT as
\begin{equation}\label{eq_spin_match}
\mO^{(\delta)}_{q\,\mu_1\ldots\mu_{\ell}}\propto
\Pi_{\mu_1\ldots\mu_{\ell}}^{\nu_1\ldots\nu_\ell}
\pd_{\nu_1}\chi\ldots\pd_{\nu_\ell}\chi(\pd\chi)^{\delta-\ell}e^{i\chi q}\,,
\end{equation}
where $\Pi_{\mu_1\ldots\mu_{\ell}}^{\nu_1\ldots\nu_\ell}$ is the projector onto traceless symmetric tensors and the overall coefficient depends on the underlying theory and operator. We omitted higher derivative contributions, which may be straightforwardly constructed as in eq. \eqref{eq_match_scalar}. Then, proceeding as before, we find that the matrix element reads
\begin{equation}
\braket{Q+q|\mO^{(\delta)}_{q\,\mu_1\ldots\mu_{\ell}}|Q}=\lambda^{(\delta,\ell)}_q\delta_{\mu_1}^0\ldots\delta_{\mu_\ell}^0\,,
\end{equation}
where $\lambda^{(\delta,\ell)}_q$ takes precisely the form in eq. \eqref{eq_OPE_threept_U(1)}. 

\subsection*{Acknowledgements}

I would like to thank Z. Komargodski, A. Monin, D. Orlando, J. Penedones, L. Rastelli and R. Rattazzi for useful discussions. My work is partially supported by the Swiss National Science Foundation under contract 200020-169696 and through the National Center of Competence in Research SwissMAP.

\bibliographystyle{JHEP}
\bibliography{biblio}{}

\end{document}